\documentclass[conference]{IEEEtran}
\IEEEoverridecommandlockouts
\usepackage{cite}
\usepackage{amsmath,amssymb,amsfonts}
\usepackage{algorithmic}
\usepackage{graphicx}
\usepackage{textcomp}
\usepackage{xcolor}
\usepackage{hyperref}
\def\BibTeX{{\rm B\kern-.05em{\sc i\kern-.025em b}\kern-.08em
    T\kern-.1667em\lower.7ex\hbox{E}\kern-.125emX}}
\begin{document}

\title{A simple protocol to automate the executing, scaling, and reconfiguration of Cloud-Native Apps}

 \author{\IEEEauthorblockN{Stanislaw Ambroszkiewicz and Waldemar Bartyna}\IEEEauthorblockA{\textit{ University of Siedlce } \\ Siedlce, Poland,\\ sambrosz@gmail.com, ORCID 0000-0002-8478-6703, waldemar.bartyna@uws.edu.pl}}

\maketitle

\begin{abstract}
The protocol specification consists of the formats of messages, and the actions taken by senders and recipients. 
The idea is that microservices of Cloud-Native Application should be also involved in configurations of their communication sessions. It does not interfere with the business logic of the microservices and requires only minor and generic modifications of the microservices codebase, limited only to network connections.  
Thus, sidecars are not needed, which is in line with the current trends, e.g.  Cilium Service Mesh.  
The formal specification of the proposed protocol is also on GitHub \cite{GitHub}, where a prototype implementation of SSMMP for a social media CNApp is also presented. The implementation clearly shows that SSMMP should be viewed (by developers) as an integral part of CNApps.  
\end{abstract}



\section{Introduction } 
This work is a continuation of \cite{Functionals-in-the-clouds}, our research into 
 the foundations of the Cloud-Native paradigm.

Microservices (as a software architecture) were first developed from the service-oriented architecture (SOA) and the concept of Web services (HTTP and  WSDL)  by Amazon  in the early 2000s. 
Hence the name AWS, which is short for Amazon Web Services.
Perhaps Amazon didn’t invent microservices alone. 
However, AWS became the most successful application of the microservices for Cloud computing at that time.

Microservice architecture comprises services that are fine-grained and protocols that are lightweight. The architecture inherited the HTTP protocol (in the form of REST) from Web services as the basic means of data transport between microservices.
Cloud Native Application (CNApp) is a distributed application composed of microservices and deployed in the Cloud.

RESTful API  is a gate for transferring data between microservices. 
The meaning (business logic) of actual communication protocols between microservices is hard-coded into the microservices. 
Therefore, when deploying a CNApp, no special requirements are required for the microservices codebase, except for this: each microservice must also be an HTTP server, i.e. it must contain, as part of its codebase, a webserver. 

To run a microservice, we need to start with a HTTP server dedicated only to that microservice.
The server must be running all the time, even if there are no client requests. The microservice must also run all the time, even when not needed.
HTTP is an application layer protocol, implemented over TCP, except HTTP/3 which is over UDP. 
What is special and unique about HTTP that it must be used to transfer data between microservices? Why can't the transfer be performed on raw TCP?
 
HTTP is the key component of the Web.
A CNApp has usually one interface to the Web; it is API Gateway. 
It seems that there is no reason to maintain the Web structure behind API Gateway and inside the Cloud cluster, especially for data transfer between microservices.
This view is gaining more and more attention, see for example
Butcher 2022 \cite{rethinking-microservices}. 

Microservices constitute an architectural pattern where a complex and sophisticated application (CNApp) is made up of a collection of fine-grained, self-contained microservices that are developed and deployed independently of each other.
They communicate over the network using protocols in accordance with the business logic of the application.

Once a collection of such microservices is composed and orchestrated into a dynamic workflow, it can be deployed on a cloud infrastructure. 

Contemporary CNApps, developed and deployed by Big Tech, consist of thousands of microservices. For example,  Uber \cite{Uber}: {\it 
``Each and every week, Uber’s 4,500 stateless microservices are deployed more than 100,000 times by 4,000 engineers and many autonomous systems. These services are developed, deployed, and operated by hundreds of individual teams working independently across the globe. The services vary in size, shape, and functionality; some are small and used for internal operations, and some are large and used for massive, real-time computation.'' } 

The scale and complexity of CNApps force the transformation of microservices from an architectural style to an organizational style, see  
Ibryam  and  Losio 2024 \cite{cloud-computing-post-serverless-trends}. It is called 
hyperspecialization of cloud services.  {\em ``
A microservice will no longer be just a single deployment unit or process boundary but a composition of functions, containers, and cloud constructs, all implemented and glued together in a single language chosen by the developer. 
The future is shaping to be hyperspecialized and focused on the developer-first cloud.'' }

Hence, a contemporary challenge in IT is to automate the deployment and management of huge and complex CNApps. This very automation is supposed to be designed by developers.

\subsection{The problem and related work} 

How to automate the executing, scaling and reconfiguration of Cloud-Native Apps in a general way, but not at the software level? 
Following Mulligan 2023 \cite{Mulligan},  this automation can be accomplished by implementing a generic protocol that extends the networking stack, on the top of TCP/IP. 

The solution we propose is the Simple Service Mesh Management Protocol (SSMMP) as a specification to be implemented in a Cloud cluster.
The specification consists of the formats of messages exchanged between the parties (actors) to the conversation of the protocol, and the actions taken by the senders and receivers of the messages.
The actors are: Manager, agents (residing on the nodes that make up the cluster), and instances of microservices running on these nodes. 
All these actors are almost the same as in Kubernetes clusters. The main difference is the abstract architecture of CNApps (introduced in Section \ref{Microservices} below). and simple 
general rules allowing for the automation of CNApps management.

Let's take a brief look at the current work on this topic.
Service Mesh is an infrastructure for CNApps that allows to transparently add security, observability and management of network traffic between the microservices without interfering with the codebase of the microservices. 
Usually, Service Mesh is built on the top of  Kubernetes and Docker. 
For an extensive overview, see, e.g.  
{\em Service Mesh Comparison} \cite{ref1-label}, and 
{\em 8 Best Service Mesh Managers to Build Modern Applications} \cite{ref2-label}. 
%
%
%
%
Each microservice is equipped with its own local proxy (called sidecar).  Sidecars can be automatically injected into Kubernetes pods, and can transparently capture all microservice traffic. The sidecars form the data plane of Service Mesh. 

The  control plane of Service Mesh is (logically) one manager responsible for configuring all proxies in the data plane to route traffic between microservices and load balancing, and to provide resiliency and security.   

Linkerd \cite{Linkerd} and Istio \cite{Istio}, both extending Kubernetes, are the best known and most popular open source software platforms for realizing  Service Mesh.
Istio uses Envoy's proxy \cite{Envoy}, while Linkerd uses its own specialized micro-proxies.

Cilium \cite{Cilium} 
 is also an open source software platform for cloud native environments such as Kubernetes clusters. It is claimed that by exploring and applying eBPF (a new revolutionary Linux kernel technology) and  WebAssembly, Cilium can challenge Docker and Kubernetes, see \cite{cni-benchmark} and \cite{webassembly-vs-container}. 
Envoy proxies are not necessary as eBPF in the kernel can replace them. 
Istio Ambient Mesh (see \cite{Betts} and \cite{Pariseau}) also follows this idea. 

While all modern Service Meshes are on the open source software level, the recent idea (see. e.g.  Mulligan 2023 \cite{Mulligan}), that the service mesh is now becoming part of the networking stack, is extremely interesting. 
It should be emphasized that the networking stack is primarily based on protocol specifications, not software. 

Open Application Model \cite{oam-dev} is ``{\em a 
platform-agnostic open source documentation that defines cloud native applications.
OAM is a new layer (abstraction) on top of Kubernetes.  
Designed to solve how distributed apps are composed and transferred to those responsible for operating them. 
Focused on application rather than container or orchestrator, Open Application Model brings modular, extensible, and portable design for defining application deployment with higher level API.}''  
While interesting in its intent, it's still just an idea suggesting that operational behaviors of CNApp need to be a part of its definition, independent of its deployment.  
A modern application should include management, monitoring, and observability components.
Moreover, the behavior should be defined in the codebase of CNApp by developer, see also Toffetti et al. 2017 \cite{toffetti2017self}. 

The topic of self-management of CNApps in service mesh has been studied for quite a long time.
Before the rise of the Cloud, it was called management of component-based distributed systems. 
There are many interesting and important works in the literature on this subject, e.g.
Di Cosmo et al. 2014 
\cite{di2014automated}, 
 Dur{\'a}n and Sala{\"u}n 2016 
 \cite{duran2016robust}, 
Toffetti et al. 2017 
\cite{toffetti2017self}, 
Etchevers et al. 2017 
\cite{etchevers2017reliable}, 
Brogi et al. 2018 
\cite{brogi2018fault}, 
Brogi et al. 2019 \cite{brogi2019robust}, 
%
\cite{kosinska2020autonomic},   
%
%
Hadded et al. 2022 
\cite{hadded2022optimal},  
Brogi et al. 2022 
\cite{brogi2022self}, 
%
%
%
Alboqmi et al. 2022 \cite{Alboqmi2022TowardES}. 
%
%
To complete the short review, also NGINX Modern Reference Architectures \url{https://github.com/nginxinc/kic-reference-architectures/} should be mentioned. 
It's an interesting idea, but still far from formal specifications.

Let's present the idea of our SSMMP.   
There are no sidecars and no proxies. 
Each microservice instance communicates (according to SSMMP) directly with the agent running on the same host.

Execution of microservices, their replications and closing are controlled  and monitored by Manager via its agents. A similar idea is also in Dur{\'a}n and Sala{\"u}n 2016 \cite{duran2016robust}. 
Communication sessions between microservices (determined by the CNApp business logic) are controlled and monitored by Manager  through its agents.

Each communication session is (like in TCP) connection-oriented. A connection between client and server needs to be established before data can be sent during the session. The server is listening for clients. 
Dynamic management of such communication sessions is the essence of the proposed protocol.

A rough description of the protocol is provided in the next two Sections \ref{Microservices}  and \ref{Abstract multi graph of CNApp}. Then, the generic functionality of the protocol is presented in Section \ref{SSMMP}. 
Our paper \cite{SSMMP} arXiv: 4889471  provides the complete formal specification of protocol messages and the corresponding actions to be performed.
The final Section \ref{Summary} is a short summary. 

\section{Microservices} 
\label{Microservices} 

CNApp is a network application where microservices communicate with each other by exchanging messages (following CNApp's business logic) using dedicated, specific protocols implemented on top of the network protocol stack. This is usually TCP/UDP/IP.
Due to its ubiquity, HTTP, implemented on the top of TCP/IP, can also be used as a transport protocol for these messages.

Each of these protocols is based on the client-server model of communication. 
 This means that the server (as part of a running microservice on a host with a network address) is listening on a fixed port for a client that is a part of another microservice, usually running on a different host. 
 Since a  client initiates a communication session with the server, this client must know the address and port number of the server. 

A single microservice can implement and participate in many different protocols, acting as a client and/or as a server.

Thus, a microservice can be roughly defined as a collection of servers and clients of the protocols it participates in, and its own internal functionality (business logic). 

Usually, communication protocols (at application layer) are defined as more or less formal specifications independently of their implementations. 

Let {\em protocol} be denoted as two closely related parties to the conversation: the server $S$ and the client $P$ which are to be implemented on two microservices. 
Formally, let protocol be denoted $(P,S)$ with appropriate superscripts and/or subscripts if needed.
After implementation, they are integral parts of microservices that communicate using this protocol.

{\em Abstract inputs} of a microservice can be defined as a collection of the servers (of the protocols) it implements:   
$$
IN:=(S_1 ,  S_2 ,  \dots   S_k)
$$ 

{\em Abstract outputs} of a microservice is defined as a collection of the clients (of the protocols) it implements:  
$$
OUT:=(P'_1 ,  P'_2 , \dots   P'_n)
$$ 

To omit confusions, the server part and a client part of a protocol will be renamed. 
Components of abstract input will be called {\em abstract sockets}, whereas  components of abstract output will be called {\em abstract plugs}. 

An abstract plug (of one microservice) can be associated to an abstract socket (of another microservice) if they are two complementary parties of the same communication protocol.   
There can be multiple abstract plugs into the same abstract socket. 

Fig. \ref{abstract-graph} presents a directed acyclic graph representing a workflow of microservices that comprise a simple CNApp. The edges of the graph are of the form (abstract plug $\to $ abstract socket). They are directed, which means that a client (of a protocol) can initiate a communication session with a server of the same protocol. 
These directions do not necessarily correspond to the data flow. This means that if a communication session is established, data (protocol messages) can also flow in the opposite direction, i.e. from an abstract input (abstract socket) to an abstract output (abstract plug).

Let us formalize the concept described above. 
{\em Microservice}  is defined as 
$$
A := (IN, \mathcal{F}, OUT)
$$
where $IN$ is the abstract inputs of the microservice, $OUT$ is the abstract outputs,  and $\mathcal{F}$ denotes the business logic of the microservice. 
%
Incoming messages, via abstract sockets of $IN$ or/and via abstract plugs of $OUT$,  invoke (as events) functions that comprise the internal functionality $\mathcal{F}$ of the microservice. This results in outgoing messages sent via $IN$ or/and $OUT$.  


The proposed  definition of microservice is at much more higher level of abstraction than TOSCA \cite{TOSCA}, an OASIS standard. 


Generally, we distinguish three kinds of such microservices. 
\begin{enumerate} 
\item
The first one is for API Gateways. They are entry points of CNApp for users.
Usually, $IN$ of API Gateway has only one element.  Its  functionality comprises in forwarding users requests to appropriate microservices. Therefore, API Gateway is supposed to be stateless.
\item 
The second kind consists of regular microservices. Their $IN$ and $OUT$ are not empty. These microservices are also supposed to be stateless. Persistent data (states) of these microservices should be stored in backend storage services (BaaS).

\item
The third kind is for backend storage services (BaaS) where all data and files of CNApp are stored. Their $OUT$ is empty. 
\end{enumerate}
From now on, all of them are also called  {\em services} of CNApp. 
%
%
Fig. \ref{abstract-graph} illustrates a CNApp composed of one API Gateway, five stateless regular microservices, and two backend storage services (BaaS).  
Note that the edges denote abstract connections and can also be seen as abstract compositions of services within a workflow.

\section{Abstract architecture }
\label{Abstract multi graph of CNApp}
\begin{figure}
  \begin{centering}
    \includegraphics[width=0.69\linewidth]{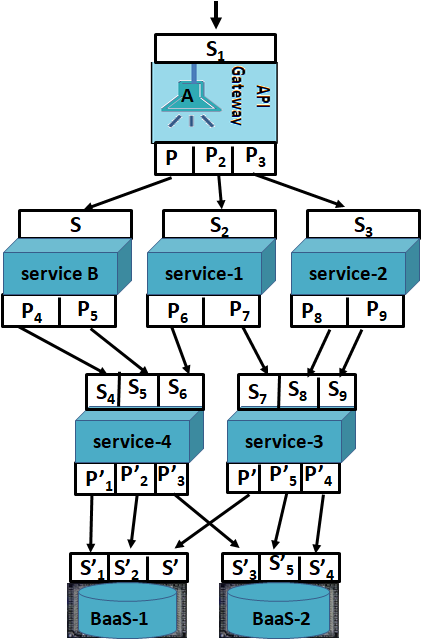}
    \caption{Abstract graph of CNApp - a simple example }
    \label{abstract-graph}
  \end{centering}
\end{figure}
{\em Abstract graph of CNApp} is defined as the following directed labeled multi-graph. 
$$
\mathcal{G}:=( \mathcal{V} , \mathcal{E} )
$$  
where $\mathcal{V}$ and $\mathcal{E}$ denote respectively Vertices and Edges. 
\begin{itemize}
\item Vertices $\mathcal{V}$ is a collection of names of services of CNApp, i.e. elements denoted in Fig. \ref{abstract-graph} as:  A (the API Gateway); regular microservices: service B, service-1, service-2, service-3, and service-4; and BaaS services: BaaS-1 and BaaS-2.   
\item 
Edges $\mathcal{E}$ is a collection of labeled edges of the graph. Each edge is of the form:  
$$
(C,\ (P,S),\ D)
$$ 
where $C$ and $D$ belong to $\mathcal{V}$,  and $(P,S)$ denotes a protocol. 
That is,  $P$ belongs to $OUT$ of $C$, and $S$ belongs to $IN$ of $D$. Hence, the edges correspond to {\em abstract connections} between microservices. The direction of an edge represents the client-server order of establishing a concrete connection.  There may be multiple edges (abstract connections) between two vertices. 
\end{itemize}
%

The above graph is an abstract view of a CNApp. Vertex  is a service name, whereas an edge is an abstract connection consisting of names of two services and the name of a communication protocol between them.  

An implementation of abstract connection $(C,\ (P,S),\ D)$ in a running CNApp may result in a concrete plug (in an instance of service $C$)  corresponding to this abstract plug $P$. The concrete plugs is connected to a concrete socket (corresponding to abstract socket $S$) of an instance of service $D$. 
This implemented connection is called a communication session and will be explained in the next Section. 


%

Initial vertices of the abstract graph correspond to API Gateways (entry points for users), whereas the terminal vertices correspond to backend storage services (BaaS) where all data and files of the CNApp are stored. 

The vertices representing regular microservices are between the API gateways and the backend storage services (BaaS).

Scaling through replication and reduction (closing replicas) of a service forces it to be stateless. The reason is that if the service is statefull, then closing (crashing) a  replica causes it to lose its state. 
We assume that API Gateways and regular microservices are stateless and can be replicated, i.e. multiple instances of such a service can run simultaneously. 


%
\begin{figure*}
  \begin{centering}
    \includegraphics[width=0.99\linewidth]{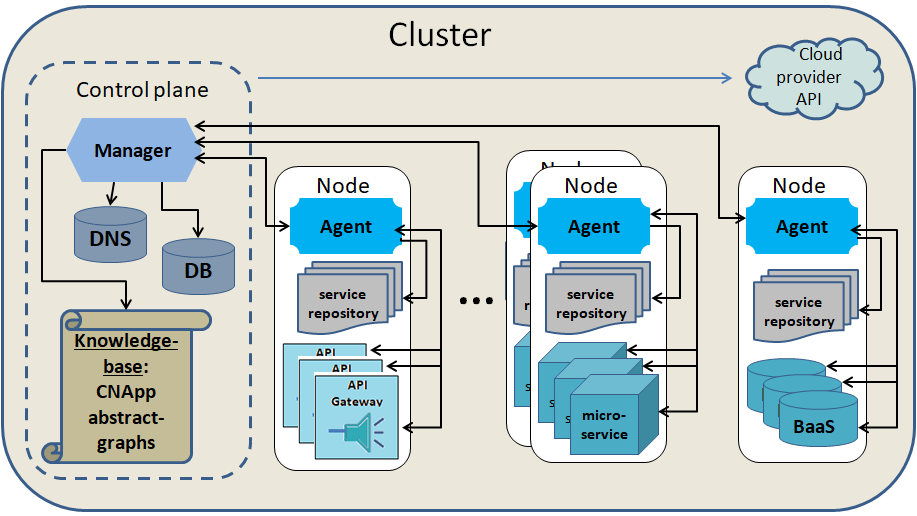}
    \caption{Simple protocol to automate the executing, scaling, and reconfiguration of Cloud-Native Apps  }
    \label{protocol}
  \end{centering}
\end{figure*}

To run CNApp, instances of its services must first be executed, then abstract connections can be configured and established as real connections, and finally protocol sessions (corresponding to these connections) can be started. 

Some services and/or connections may not be used by some  executions of CNApp. Temporary protocol sessions can be started for already established connections (and then closed along with their connections) dynamically at runtime. Multiple service instances may be running, and some are shutting down.
This requires dynamic configurations of network addresses and port numbers for plugs and sockets of the instances. 
The novelty of SSMMP lies in the smart use of these configurations. A similar idea has been used by Netflix \cite{Netflix} at the software level. 


\section{Simple service mesh management protocol - SSMMP} 
\label{SSMMP}

The complete formal specification of SSMMP is in \cite{SSMMP}. 
Here we will present the protocol in an intuitive, somewhat informal way.
%
%
The main actors of the protocol are: Manager, agents, and running instances of services (API Gateways, regular microservices, and BaaS services), see Fig. \ref{protocol}.  

There may be two (or more) running instances of the same service. Hence, the term service refers rather to its bytecode. 
%

Manager communicates only with the agents. 
Agent, on a node, communicates with all service instances running on that node. 
Within the framework of SSMMP,  any service instance (running on a node) can only communicate with its agent on that node. 

Agent has a service repository at its disposal. It consists of bytecodes of services that can be executed (as service instances) on this node by the agent. The agent (as an application) should have operating system privileges to execute applications and to kill application processes.
Agent acts as an intermediary in performing the tasks assigned by the Manager.
All service instance executions as well as shutting down running instances are controlled by the Manager through its agents.

Each agent must register with Manager so that the network address of its node and its service repository are known to Manager.

 

At a request of Manager, agent can execute instances of services whose bytecodes are available in its repository or shut down these instances. 

Once a service instance is executed, it initiates the SSMMP communication session with its agent.
The network address of the agent is, of course, {\small\tt localhost} for the all service instances running on the same node (host). The port number of the agent (to communicate with its service instances) is fixed for SSMMP, and is the same for all the agents. 

The agent can monitor the functioning of service instances running on its node (in particular, their communication sessions) and report their status to Manager.

Manager can also shut down (via its agent) a running instance that is not being used, is malfunctioning, or is being moved to another node.

Usually, in the existing service meshes, Manager controls the execution of CNApps in accordance with a policy defined by the Cloud provider. 

In SSMMP, the design and implementation of an instance of Manager is delegated to the developer of CNApp, who can take into account the cloud provider's policies. This makes this instance (dedicated to this CNApp) an integral part of the CNApp.
Current state of Manager as well as its history are stored in a dedicated database DB.  Manager knows the service repositories of all its agents. 

The Knowledge-base of Manager consists of abstract graphs of CNApps, i.e. the CNApps that can be deployed on the cluster comprising all the nodes.   

The current state of any running instance of service is stored in Manager's database, and consists of:
\begin{enumerate}
\item
		open communication sessions and their load metrics; 
\item
	 	observable (healthy, performance and security) metrics, logs and traces.  
\end{enumerate}
%


%
The key element of SSMMP is the concept of  {\em communication session}  understood jointly as establishing a connection and then starting a protocol session on this connection.%
The process of establishing and closing such sessions is controlled by the Manager through its agents. This is explained in detail below. 

Communication session is an implementation of an abstract  connection, say $(A, (P,S), B)$ that is an edge of the abstract graph of CNApp; Fig. \ref{abstract-graph} may serve as an example where $A$ is API Gateway, and $B$ is service $B$. 
The service name $B$ (as a parameter of the abstract connection) is not encoded explicitly in service $A$. It must be given by Manager as a configuration parameter (according to the abstract graph) for execution of an instance of service $A$ by an agent. 
Thus, service $A$  as well as the all services are supposed to be generic, i.e. $A$ may be used as a component of another CNApp for a connection, say $(A, (P,S), C)$,  where $C$ is different than $B$. 

In order to execute an instance of the service $A$, the Manager sends a request to an agent (which has the  bytecode of $A$ in its repository, and resides on a node with a fixed network address) to execute that bytecode. The request has parameters for this execution, which include: ports numbers for the sockets of $A$ and service names for its plugs. 
 For the plug $P$ it is $B$. 

Let an instance of service $A$ (denoted $i$) be running. 
%
To implement $P$ (as a client) in the instance $i$, this instance needs to have  a translation of  the parameter $B$ to the network address of the host where an instance $j$ of the service $B$ is already running, and the port number on which the socket $S$ (as a server) of the instance $j$ listens to clients.

The port number of this socket, and generally the ports numbers of all  sockets of $B$, are configured by Manager as parameters dedicated to that very instance execution. 
This allows multiple instances of the same service to run on the same node (same network address but different port numbers for the sockets).

In order to implement the abstract connection  $(A, (P,S), B)$, the  instance $i$ sends a request to Manager (via its agent) to translate the parameter $B$. The Manager responses (via the agent) with a translation. 
The translation contains the network address of a running instance of $B$, and the port number of socket $S$ of the instance. Then, a  communication session (as an implementation of the abstract connection  $(A, (P,S), B)$) can be established by the instance $i$. 

If a service instance is running and not in use, there was no reason to start it. Therefore, service instances should be started only when they are needed and shut down when they are no longer needed.
The same applies to establishing communication sessions. These two aspects are closely related, i.e. if all sessions of a running service instance (which is not an API Gateway) are closed or have not been established  for some predetermined period of time, then the instance should be shut down. 

{\em Current state} of a running CNApp is defined by running service instances, and already established (and not closed) communication sessions between the instances. 
%
%
%
It is the basis for management decisions made by Manager. 
There are four kinds of such decisions: 
\begin{itemize}
\item 
execution and shutdown of service instances, 
\item 
load balancing by multiple instance executions of stateless services, and closing some of them, 
\item establishing or closing  communication sessions,   
\item
and reconfiguration of running instances; this includes moving some instances to other nodes.  
\end{itemize}
These decisions (mutually interrelated) are forwarded to appropriate agents as tasks to be accomplished. 

Usually, API Gateway is the Web interface for users of CNApp.  
In Fig. \ref{abstract-graph}, socket $S_1$ implements  HTTP server listening on default port number 80. 
Multiple instances of an API Gateway can be executed (on the basis of  DNS load balancing controlled by Manager) so that users requests are distributed across many   instances of the API Gateway. 
This is done in the following way. 
The alias (the name) of an API Gateway and the port number (by default it is 80)  are  supposed to be known to all users of the CNApp. 
Execution of an instance of the API Gateway is done in the following way. 
\begin{enumerate}
\item 
Manager sends a request to an agent residing on a node to execute an instance of the API Gateway. The request includes a configuration of the plugs of that API Gateway. In Fig. \ref{abstract-graph} the plugs are $P$, $P2$, and $P_3$. 
%
It is supposed that the bytecode of that API Gateway belongs to the agent's repository. 
\item
The agent  executes the instance, and sends the confirmation to Manager. 
\item 
Manager stores the  instance identifier (as a canonical name) and its network address in its database, and adds two records to its DNS:  a record of type A, and  a record of type CNAME. 
Thus, a request to resolve the API Gateway name (alias) via DNS is answered by sending the network address of one of the running API Gateway instances.
\end{enumerate}

Note that no fixed port numbers for sockets are a priori assigned to any regular service or any BaaS service.
The exceptions are API Gateways, where port numbers are fixed and should be well known to users.

Execution of an instance of a regular service or a backend storage service is done as follows. Let us consider $service-1$ from Fig. \ref{abstract-graph} as an example. 
\begin{enumerate}
\item 
Manager sends a request to an agent residing on a node to execute an instance of ${service-1}$. It is supposed that 
the bytecode of ${service-1}$ belongs to the agent's repository. The request includes the configuration of the port numbers  (assigned by Manager) for all sockets of ${service-1}$. In this case, it is one socket $S_2$. The configuration of the plugs of ${service-1}$ (i.e. plugs $P_6$ and $P_7$) consists in assigned to each such plug a service name according the the abstract graph of the CNApp. 
In Fig. \ref{abstract-graph}, plug $P_6$ is assigned to  ${service-4}$  according to the edge $({service-1}, (P_6, S_6),  {service-4})$. Similarly, the plug $P_7$ is assigned to ${service-3}$ according to the edge \\ 
$({service-1}, (P_7, S_7),  {service-3})$. 
\item
The agent executes an instance of ${service-1}$  for these configurations.  
\item 
Agent sends to Manager the confirmation of the successful execution. 
\item 
Manager stores in its database DB the following items: the name of the service, identifier of the instance, the network address of the instance,  the configuration of port numbers of the sockets, and the configuration of the plugs. This is crucial for establishing new communication sessions that this instance will participate in.
\end{enumerate}

Establishing a communication session for abstract connection $({service-1}, (P_6, S_6),  {service-4})$ is done in the following way. 
\begin{enumerate}
\item 
An already  running instance $i$ of  ${service-1}$ sends a request to its agent for: the network address of a node where an instance of ${service-4}$ is running, and the port number of the socket $S_6$ of this instance. 
\item
The request is forwarded to Manager that is responsible either to choose (from its DB) a node where an instance of ${service-4}$ is already running, or to execute new instance of service ${service-4}$  on a node. 
\item 
Manager sends the network address and the port number of $S_6$ of a running instance (denoted  $j$) of ${service-4}$ to the agent. Then, the agent forwards it to instance $i$ of ${service-1}$ where a concrete plug for the abstract plug $P_6$ can be construed now, and the communication session can be established to socket $S_6$ of $j$. 
\end{enumerate}

Closing an existing communication session (between running instance $i$ of  $C$ and a running instance $j$ of  $D$)  for abstract connection $(C, (P,S), D)$ can be initiated by one of the instances, like for the TCP connection. Then, the services must report the successful closing to their agents. These reports are forwarded to Manager. 

The session closing may be also requested by Manager via the agents. The agents forward the request to the instances. The instances close the session, and report this to their agents. The agents forward these reports to Manager. 

The Manager can request the agent to shut down a running service instance.
For graceful shutdown of a running instance, all its  communication sessions should be closed beforehand. Then, an internal method (like {\small\tt System.exit()} in Java) can be evoked to shut down the instance. 

Manager can enforce (via its agent) a hard shutdown of a running instance of a service by killing the instance process. This can be done also when a running instance fails, e.g. does not respond to its agent or is malfunctioning. 
Before this hard shutdown, all communication sessions in which this instance was  participated should be closed by participants of opposing sides of the session at the request of the Manager through its agents.

If that shutdown instance was for an API Gateway, Manager removes the records (related to that instance) from its DNS.

%
%


%

\section{Summary} 
\label{Summary}

SSMMP is simple if we consider its description, and especially the complete formal specification presented below. Its implementation (as a proof of concept) is at GitHub \cite{GitHub} with  the complete Java API \url{https://github.com/sambrosz/SSMMP-a-simple-protocol-for-Service-Mesh-management/tree/main/ssmmpComAPIv1.1.pdf}.  

The concept of abstract connection between services (in the abstract graph of CNApp) and its implementation as communication sessions is crucial. 
The  abstract definition of service of CNApp  is also important here. 
Separation of these abstract notions from deployment is important. 

The novelty of SSMMP consists in the dynamic establishment and closing of communication sessions at runtime based on the configurations assigned to sockets and plugs by the Manager.

Although a similar approach has already been used in Netflix \cite{Netflix} (as dedicated software), it can be fully exploited in Netflix by extending the network protocol stack with SSMMP.

Since executing, scaling and reconfiguration of CNApp can be done by SSMMP,
it seems reasonable to include SSMMP as an integral part of CNApp. 
Then, also CNApp crash recovery could be performed via SSMMP.
%
%

Graph of CNApp and the states of its running instances are stored by Manager in its  KB and DB. 
Failures of agents and service instances can be handled if Manager is running properly. 
Replications of cluster nodes, agents and their service repositories are sufficient means for recoveries from such failures. 

The central Manager is the weakest point here; its failure results in an irreversible failure of the running CNApp.
However, if the Manager's current state is kept securely by a supervising manager, the Manager process can be recovered from that state. 
The supervising manager can also serve as a distributed control plane where there are several Managers, each controlling a portion of the CNAApp abstract graph.
Netflix uses over 1000 microservices now. Uber now has 4500 or more independent microservices.  A distributed control plane is necessary for such huge CNApps. 

 SSMMP was designed to be independent of transport and network protocol stack. TCP/IP is the default stack for communication sessions. Named Data Networking may be seen as an alternative. 

\section{APPENDIX: SSMMP/1.1 - specification of the protocol messages and actions}
\label{SSMMP}  
From now on, we will use normal (not italic) letters to denote services, their instances, plugs, sockets and connections, like  A, P, S, B,  i, and  j. 

There are two general kids of messages: request, and response to this request. 
All messages are strings. 
Message consists of a sequence of lines. 
Line is of the form: 
{\small \begin{verbatim}
line_name: contents
\end{verbatim}}

The first line is for message type. 
The second line is for message identifier (an integer). The identifier is unique, and is the same for request and its response. 

\subsection{Initialization of the protocol}

Let a CNApp be fixed. The abstract graph of CNApp is known to Manager. For each service of CNApp, there are agents that can execute instances of this service, i.e. the bytecode of this service belongs to the agents repositories. 

Each entry of the repository is of the form (service-name, list of socket names, list of plug names, bytecode of service).  

An agent registers with Manager and sends the list of service names of its repository. 
Request for the registration from agent to Manager is of the following form. 

{\small \begin{verbatim}
type: initiation_request 
message_id: [integer]
agent_network_address: [IPv6]
service_repository: [service name list]
\end{verbatim}}
In square brackets of {\small \tt message\_id: [integer]} there is an element of a datatype, in this case it is a positive integer determined by the agent. 
 In the case of {\small \tt agent\_network\_address: [IPv6]}, it is a concrete IPv6 network address.  
For the line {\small \tt service\_repository: [service name list]}, the contents  denotes a sequence of the  form {\small \tt (name\_1; name\_2; \dots name\_k)}. 

Registration response form Manager to agent is as follows.  

{\small \begin{verbatim}
type: initiation_response 
message_id: [integer]
status: [status code]
\end{verbatim}}
Universal HTTP response status codes are proposed to be adapted to SSMMP. Their meaning depends on response types. Each code consists of three digits, and is of the form:
\\
 {\small \tt 1xx  informational response}, \\
 {\small \tt 2xx successful} – the request was successfully received, understood, and accepted,\\
 {\small \tt 3xx redirection} – further action needs to be taken in order to complete the request,\\
 {\small \tt 4xx requester error} – the request contains bad syntax or cannot be fulfilled,\\
 {\small \tt 5xx respondent error}.

\subsection{Execution of service A} 

 Manager assigns a unique identifier, say i, (a positive integer) to a new instance of A to be executed. Manager also determines port numbers to all sockets of the instance.  It is called socket configuration, and is a sequence of pairs: 
{\small \begin{verbatim}(socket_name, port_number)\end{verbatim}}
 
Configuration of plugs is a sequence of pairs: 
{\small \begin{verbatim}(plug_name, service_name)\end{verbatim}} assigned to instance i by Manager according to the CNApp abstract graph. This means that each abstract plug is assigned a service name, where is the corresponding abstract socket.
Manager also determines a network address (denoted {\small\tt NA\_i}) of the node where the instance i of A is to be executed by the agent residing on that node. 

Request from Manager to the agent to execute the instance i of service A is as follows. 
{\small \begin{verbatim}
type: execution_request 
message_id: n
agent_network_address: NA_i
service_name: A
service_instance_id: i
socket_configuration: [configuration of sockets]
plug_configuration: [configuration of plugs]
\end{verbatim}}

Action of the agent:  execution of instance i of the service A for these  configurations of sockets and plugs. 

Response from agent to Manager: 

{\small \begin{verbatim}
type: execution_response 
message_id: n
status: [status code]
\end{verbatim}}
%

\subsection{Communication session establishment}
\label{Communication session establishment}
\noindent 
Establishing a communication session for abstract connection $(A, (P,S), B)$  between  instance i of A, and instance j of B.  

Let us suppose that instance i of service A is already running on the node that has network address {\small\tt NA\_i}. 

Request from the instance i of service A to its agent:
 {\small \begin{verbatim}
type: session_request 
message_id: n
sub_type: service_to_agent
source_service_name: A
source_service_instance_id: i
source_service_instance_network_address: NA_i
source_plug_name:  P
dest_service_name: B
dest_socket_name:  S
\end{verbatim}} 

Request is forwarded to Manager by the agent:  

{\small \begin{verbatim}
type: session_request 
message_id: n
sub_type: agent_to_Manager
agent_network_address: NA_i
source_service_name: A
source_service_instance_id: i
source_service_instance_network_address: NA_i
source_plug_name:  P
dest_service_name: B
dest_socket_name:  S
\end{verbatim}}

If there is no instance of service B already running, then 
 Manager sends a request to an agent to execute an instance j  of service B. 
Otherwise, i.e. if instance j of service B (on the node with  network address {\small\tt NA\_j} and the port k for S) is already running, then Manager sends the following response to the agent: 
{\small \begin{verbatim}
type: session_response 
message_id: n
sub_type: Manager_to_agent
source_service_name: A
source_service_instance_id: i
dest_service_name: B
dest_service_instance_id: j 
dest_socket_name: S
dest_service_instance_network_address: NA_j
dest_socket_port: k
status: [status code]
\end{verbatim}}

Response from agent to instance of A: 
{\small \begin{verbatim}
type: session_response 
message_id: n
sub_type: agent_to_service
source_service_name: A
source_service_instance_id: i
dest_service_name: B
dest_service_instance_id: j 
dest_socket_name: S
dest_service_instance_network_address: NA_j
dest_socket_port: k
status: [status code]
\end{verbatim}} 
Action of instance i of A: initialize  $(P,S)$ session to instance j of B. The port number of plug P is determined; let it be denoted by m. 

By default, the socket S of instance j of B accepts the session establishment.  This acceptance will  be known to Manager, if the session acknowledgment is send by instance i to Manager via the agent. 

Action of instance j of B: accept the establishing $(P,S)$ session to instance i of A. 
New socket port number (say l) is assigned to this session; this is exactly the same as for  TCP connection.  

Instance j of B gets to know the values of the parameters: 
{\small \begin{verbatim}
source_service_instance_network_address: NA_i
source_plug_port: m
\end{verbatim}}
 Instance i of A gets to know the value of the parameter {\small\tt dest\_socket\_new\_port: l}
 \\ 
Acknowledgment of the  established session is sent  by the instance i of  A to its agent.  

{\small \begin{verbatim}
type: session_ack
message_id: n
sub_type: service_to_agent
source_service_name: A
source_service_instance_id: i
source_plug_port: m
dest_socket_new_port: l
status: [status code]
\end{verbatim}} 

and forwarded to Manager by the agent: 
{\small \begin{verbatim}
type: session_ack
message_id: n
sub_type: agent_to_Manager
source_service_name: A
source_service_instance_id: i
source_plug_port: m
dest_socket_new_port: l
status: [status code]
\end{verbatim}}

Note that  {\small\tt message\_id} is n (determined by instance i of A), and is the same for all the above request, response and acknowledgment messages. 


The complete list of the parameters of the session is as follows. 
{\small 
\begin{verbatim}
source_service_name: A
source_service_instance_network_address: NA_i
source_service_instance_id: i
source_plug_name: P
source_plug_port: m
dest_service_name: B 
dest_service_instance_network_address: NA_j
dest_service_instance_id: j
dest_socket_name: S
dest_socket_port: k
dest_socket_new_port: l
\end{verbatim}
}
The number k is the port number (assigned by Manager) to the socket S (of instance j of B) for listening to clients. 
New port l of the socket S is dynamically assigned by instance j of B solely for the communication session with P of instance i of A. 

The instance i of A knows the above parameters except {\small\tt dest\_service\_instance\_id: j}. 

The instance j of B knows the above parameters except {\small\tt source\_service\_instance\_id: i \\
source\_service\_name: A
}

Manager knows all parameters of the session.

\subsection{Closing a communication session }

Suppose that there is an already established session of the protocol $(P,S)$ between running instance i of A and instance j of B. 

Each of the instances can initialize session closing, like in TCP connection, according to its own business logic. 
If instance i does so, it informs instance j of B that does the same; and vice versa. 
This is a regular closing of the session. 

A session may be closed only by one part of the communication due to failure of the other part or a broken link making the communication between these two parts impossible. 
In any of these cases above a running instance sends a message to its agent informing that the session was closed. 
Then, the agent forwards it to Manager. 

\subsubsection{Session closing resulting from business logic by instance i of A}
The message from instance i of A to its agent is as follows. 

{\small \begin{verbatim}
type: source_service_session_close_info
message_id: n
sub_type: source_service_to_agent
source_service_name: A
source_service_instance_id: i
source_service_instance_network_address: NA_i
source_plug_name: P
source_plug_port: m
dest_service_name: B
dest_service_instance_id: j 
dest_service_instance_network_address: NA_j
dest_socket_name: S
dest_socket_port: k
dest_socket_new_port: l
status: [status code]
\end{verbatim}}
The value of status code is 111 if the closing results from business logic, or is 122 if the TCP connection is broken 

Note that the instance i of A sends all (known to it) parameters of the session. 
The agent forwards the info to Manager:
{\small \begin{verbatim}
type: source_service_session_close_info
message_id: n
sub_type: agent_to_Manager
source_service_name: A
source_service_instance_id: i
source_service_instance_network_address: NA_i
source_plug_name: P
source_plug_port: m
dest_service_name: B
dest_service_instance_id: j 
dest_service_instance_network_address: NA_j
dest_socket_name: S
dest_socket_port: k
dest_socket_new_port: l
status: [status code]

\end{verbatim}}
The value of the parameter 
{\small\tt message\_id: n} 
is the same for the both messages above, and is determined by instance i of A. 
Manager can determine  the identifier j of the instance of B on the basis of the port numbers: m, k and l. 

\subsubsection{Session closing resulting from business logic by instance j of B for connection (A, (P,S), B)} 

Session closing by instance j of B is similar. 
The message from instance j of B to its agent  is as follows. 

{\small \begin{verbatim}
type: dest_service_session_close_info
message_id: o
sub_type: dest_service_to_agent
source_service_instance_network_address: NA_i
source_plug_name: P
source_plug_port: m
dest_service_name: B
dest_service_instance_network_address: NA_j
dest_service_instance_id: j
dest_socket_name: S
dest_socket_port: k
dest_socket_new_port: l
status: [status code]
\end{verbatim}}
The instance j of B sends all (known to it) parameters of the session. 
The agent forwards the info to Manager. 
{\small \begin{verbatim}
type: dest_service_session_close_info
message_id: o
sub_type: agent_to_Manager
source_service_instance_network_address: NA_i
source_plug_name: P
source_plug_port: m
dest_service_name: B
dest_service_instance_network_address: NA_j
dest_service_instance_id: j
dest_socket_name: S
dest_socket_port: k
dest_socket_new_port: l
status: [status code]
\end{verbatim}}
The value of the parameter 
{\small\tt message\_id: o} 
is the same for both messages above and is determined by instance j of B. 
Manager can determine  the identifier i of the instance of A on the basis of the port numbers: m, k and l.

\subsubsection{Session closing on the request of Manager for connection (A, (P,S), B)}

An initiation of a communication session for abstract connection $(A, (P,S), B)$  between  instance i of A, and instance j of B is done  by the instance i according to its business logic. Upon the request of instance i, configuration for such session is sent to instance i by Manager via the agent of instance i. 

By default, the socket S of instance j of B accepts the session establishment.  This acceptance is known to Manager. 

Manager's request to close this session is only sent to the instance i of service A via the agent of instance i. 
The request contains all parameters of the session known to the instance i of A, i.e. except {\small\tt dest\_service\_instance\_id: j}. 

Request, from Manager to instance i of A via its agent to close a session, is as follows. The value  o of the parameter  {\small\tt message\_id: } is determined by Manager. 

{\small \begin{verbatim}
type: source_service_session_close_request
message_id: o
sub_type: Manager_to_agent
source_service_name: A
source_service_instance_id: i
source_service_instance_network_address: NA_i
source_plug_name: P
source_plug_port: m
dest_service_name: B
dest_service_instance_network_address: NA_j
dest_socket_name: S
dest_socket_port: k
dest_socket_new_port: l
\end{verbatim}}

The agent  forwards the request to instance i of A 

{\small \begin{verbatim}
type: source_service_session_close_request
message_id: o
sub_type: agent_to_source_service
source_service_name: A
source_service_instance_id: i
source_service_instance_network_address: NA_i
source_plug_name: P
source_plug_port: m
dest_service_name: B
dest_service_instance_network_address: NA_j
dest_socket_name: S
dest_socket_port: k
dest_socket_new_port: l
\end{verbatim}}

Action of instance i of A: closing P.  

Response of instance i of A to its agent:
{\small \begin{verbatim}
type: source_service_session_close_response
message_id: o
sub_type: source_service_to_agent
status: [status code]
\end{verbatim}}
Forwarding the response to Manager. 
{\small \begin{verbatim}
type: source_service_session_close_response
message_id: o
sub_type: agent_to_Manager
status: [status code]
\end{verbatim}}
After the successful session closing, the instance i may need a new communication session (for the same connection) to complete the task interrupted by the enforced closing. In order to do so the instance i can send a request to Manager for a configuration needed to establish such session. The message format of this request is 
given in Section \ref{Communication session establishment}. 

In the case of failure of instance i of service A, or its agent or node, a similar request must be sent to instance j of service B to close the session. This requires only minor modifications to the message sequence above.

\subsubsection{Session closing on the request of Manager for connection (C, (P,S), A)}

Request, from Manager to instance j of A via its agent to close a session, is as follows. The value o of the parameter    message\_id:  is determined by Manager.

{\small \begin{verbatim}
type: dest_service_session_close_request 
message_id: o
sub_type: Manager_to_agent 
dest_service_name: A
dest_service_instance_id: j
dest_service_instance_network_address: NA_j 
source_service_instance_network_address: NA_i 
source_plug_name: P
source_plug_port: m
dest_socket_name: S
dest_socket_port: k
dest_socket_new_port: l
\end{verbatim}}
The agent forwards the request to instance j of A 

{\small \begin{verbatim}
type: dest_service_session_close_request 
message_id: o
sub_type: agent_to_dest_service 
dest_service_name: A
dest_service_instance_id: j
dest_service_instance_network_address: NA_j 
source_service_instance_network_address: NA_i 
source_plug_name: P
source_plug_port: m
dest_socket_name: S
dest_socket_port: k
dest_socket_new_port: l
\end{verbatim}}
Action of instance j of A: closing socket on port l. 
Response of instance j of A to its agent:
{\small \begin{verbatim}
type: dest_service_session_close_response
message_id: o
sub_type: dest_service_to_agent
status: [status code]
\end{verbatim}}
Forwarding the response to Manager. 
{\small \begin{verbatim}
type: dest_service_session_close_response
message_id: o
sub_type: dest_service_to_Manager
status: [status code]
\end{verbatim}}

After the successful session closing, the instance j of service A may need a new communication session (for the same connection) to complete the task interrupted by the enforced closing. In this case, socket S must wait for a new client request.  
In the case of failure of instance j of service A, or its agent or node, a similar request must be sent to instance i of service C to close the session.

%
\subsection{Shutdown of service instance }

 Graceful shutdown of a running  instance of service by itself (on a request of Manager forwarded by agent)  can be done after closing of all its communication sessions on the request of Manager via agent. The appropriate requests and responses  are as follows. 
From Manager to agent: 
{\small \begin{verbatim}
type: graceful_shutdown_request
message_id: o
sub_type: Manager_to_agent
service_name: A
service_instance_id: i
\end{verbatim}}
From agent to service instance:
{\small \begin{verbatim}
type: graceful_shutdown_request
message_id: o
sub_type: agent_to_service_instance
service_name: A
service_instance_id: i
\end{verbatim}}
Instance i of service A invokes internal method to shut down itself. Just before completing it, the instance sends the following response to the agent. 
{\small \begin{verbatim}
type: graceful_shutdown_response
message_id: o
sub_type: service_instance_to_agent
status: [status code]
\end{verbatim}}
The agent forwards the response to Manager: 
{\small \begin{verbatim}
type: graceful_shutdown_response
message_id: o
sub_type: agent_to_Manager
status: [status code]
\end{verbatim}}
%

\subsubsection{Hard shutdown of  instance  i of service A is done by agent on the request of Manager} 

{\small \begin{verbatim}
type: hard_shutdown_request
message_id: n
sub_type: Manager_to_agent
service_name: A
service_instance_id: i
\end{verbatim}}
Action of the agent: kill the process  of service instance i. 

The response is as follows. 
{\small \begin{verbatim}
type: hard_shutdown_response
message_id: n
sub_type: agent_to_Manager
service_instance_id: i 
status: [status code]
\end{verbatim}}
%

\subsubsection{Termination of service instance resulting from business logic} 

Termination of instance i of service A 
Before System.exit() the following message is sent to Agent
{\small \begin{verbatim}
type: service_instance_termination_info
message_id: n
sub_type: service_instance_to_agent
service_name: A
service_instance_id: i
service_instance_network_address: NA_i 
\end{verbatim}}
Agent forwards the message to Manager

{\small \begin{verbatim}
type: service_instance_termination_info
message_id: n
sub_type: agent_to_Manager
service_name: A
service_instance_id: i
service_instance_network_address: NA_i 

\end{verbatim}}
%

\subsection{Simple monitoring of service instances by Manager}

Manager's request for observable metrics from a service instance is as follows. 

{\small \begin{verbatim}
type: service_instance_health_request
message_id: o
sub_type: Manager_to_agent
service_instance_id: i
service_instance_network_address: NA_i 

\end{verbatim}}
Agent forward this message to service instance: 
{\small \begin{verbatim}
type: service_instance_health_request
message_id: o
sub_type: agent_to_service_instance
service_instance_id: i
service_instance_network_address: NA_i 
\end{verbatim}}
Response of service instance : 
{\small \begin{verbatim}
type: service_instance_health_response
message_id: o
sub_type: service_instance_to_agent
service_instance_id: i
status: [status code]
\end{verbatim}}
The status codes may express the metrics. 
Agent forwards the response to Manager. 

{\small \begin{verbatim}
type: service_instance_health_response
message_id: o
sub_type: agent_to_Manager
service_instance_id: i
status: [status code]
\end{verbatim}}
%

\subsection{Simple monitoring of Agent by Manager} 

Manager's request for observable metrics from Agent
{\small \begin{verbatim}
type: agent_health_control_request
message_id: o
sub_type: Manager_to_agent
agent_network_address: NA 
\end{verbatim}}
Response from Agent to Manager
{\small \begin{verbatim}
type: agent_health_control_response
message_id: o
sub_type: agent_to_Manager
agent_network_address: NA 
status: [status code]
\end{verbatim}}
%

\subsection{Final remarks } 

Status codes can be used to handle failures. This is left to  SSMMP implementations.

Requirements for developing services of CNApp participating in SSMMP  are as follows. 
Each instance of a service of CNApp is obliged to close its communication session at the request of the Manager. This can interfere with the business logic of the instance.
For this reason, the current state of the communication session (until closed) must be stored in a BaaS service. To continue a task interrupted by the closing, the instance (at the client side of the connection) can establish a new session for the same abstract connection to continue and possibly to complete the task. Retrieval of the current state from the BaaS service may also be needed. 
This is the most complex requirement to be implemented in the codebase of each service participating in SSMMP. 
This requirement can also be seen as a standard recovery mechanism (independent of SSMMP) for handling failures of communication session, e.g. resulting from broken network connections. It seems reasonable to implement these recovery mechanisms in each CNApp service, regardless of SSMMP.
The rest of the implementation requirements are relatively simple and can be completely separated from the business logic of the services. For details,  see the complete Java API \url{https://github.com/sambrosz/SSMMP-a-simple-protocol-for-Service-Mesh-management/tree/main/ssmmpComAPIv1.1.pdf}. 
%
%



\bibliographystyle{IEEEtran}
\bibliography{FC}
\end{document}